\documentstyle[12pt]{article}
\input{epsf}
\newcommand{\be}{\begin{eqnarray}}
\newcommand{\ee}{\end{eqnarray}}
\newcommand{\ba}{\begin{array}}
\newcommand{\ea}{\end{array}}
\newcommand{\half}{{\textstyle{\frac{1}{2}}}}
\newcommand{\partialslash}{\partial\hspace{-.5em}/\hspace{.15em}}
\newcommand{\Pslash}{P\hspace{-.5em}/\hspace{.15em}}
\newcommand{\Qslash}{Q\hspace{-.5em}/\hspace{.15em}}
\newcommand{\kslash}{k\hspace{-.5em}/\hspace{.15em}}
\newcommand{\bfk}{{\bf k}}
\newcommand{\kint}{\int_{\Lambda_3}\!\frac{d^4 k}{(2\pi)^4}}
\newcommand{\fourint}[1]{\int\!\frac{d^4 #1}{(2\pi)^4}}
\begin{document}
%
%
\rightline{RUB-TPII-12/96}
\rightline{hep-ph/9608347}
\vspace{1cm}
\begin{center}
\begin{large}
{\bf A chiral Lagrangian for excited pions} \\[1.5cm]
\end{large}
{\bf M.K.\ Volkov}$^{\rm 1}$ \\[0.4cm]
{\em Bogoliubov Laboratory of Theoretical Physics \\
Joint Institute for Nuclear Research, Dubna \\
Head Post Office, P.O. Box 79, 101000 Moscow, Russia}
\\[0.7cm]
{\bf C. Weiss}$^{\rm 2}$ \\[0.4cm]
{\em Institut f\"ur Theoretische Physik II \\
Ruhr--Universit\"at Bochum \\
D-44799 Bochum, Germany}
\end{center}
\vspace{1cm}
\begin{abstract}
\noindent
We construct a chiral Lagrangian containing besides the usual pion
field ($\pi$) also its first radial excitation ($\pi'$). The
Lagrangian is derived by bosonization of a Nambu--Jona-Lasinio quark
model with separable non-local interactions, with form factors
corresponding to 3--dimensional ground and excited state wave
functions. Chiral symmetry breaking is governed by the NJL gap
equation. The effective Lagrangian for $\pi$-- and $\pi'$--mesons
shows the decoupling of the Goldstone pion and the vanishing of the
$\pi'$ leptonic decay constant, $f_{\pi'}$, in the chiral limit, as
required by axial current conservation. We derive the excited states'
contribution to the axial current of the model using Noether's
theorem. For finite pion mass and $\pi'$ masses in the range of 
$750\,{\rm MeV}$--$1300\, {\rm MeV}$, $f_{\pi'} / f_\pi$ is found to 
be of the order of 1\% .
\end{abstract}
\vspace{1cm}
\vfill
\rule{5cm}{.15mm}
\\
\noindent
{\footnotesize $^{\rm 1}$ E-mail: volkov@thsun1.jinr.dubna.su} \\
{\footnotesize $^{\rm 2}$ E-mail: weiss@hadron.tp2.ruhr-uni-bochum.de}
\newpage
%
%
%
\section{Introduction}
Radial excitations of light mesons are currently a topic of great
interest in hadronic physics. During the next years, facilities at
CEBAF and IHEP (Protvino) will provide improved experimental
information about excited states in the few--GeV region, {\em e.g.}\
on the $\pi'$ meson. The resonance usually identified as the first
radial excitation of the pion has a mass of 
$(1300 \pm 100)\,{\rm MeV}$ \cite{Rev_94}.  Recently, indications of 
a light resonance in diffractive production of $3\pi$--states have
lead to speculations that the mass of the $\pi'$ may be considerably
lower at $\sim 750 \,{\rm MeV}$ \cite{ivanshin_93}.
\par
A theoretical description of radially excited pions poses some
interesting challenges. The physics of normal pions is completely
governed by the spontaneous breaking of chiral symmetry. A convenient
way to derive the properties of soft pions is by way of an effective
Lagrangian based on a non-linear realization of chiral symmetry
\cite{CCWZ69}. When attempting to introduce higher resonances to
extend the effective Lagrangian description to higher energies, one
must ensure that the introduction of new degrees of freedom does not
spoil the low--energy theorems for pions, which are universal
consequences of chiral symmetry. In the case of vector resonances
($\rho , \omega , A_1$) this problem can be solved by introducing
them as gauge bosons. Such ``gauged'' chiral Lagrangians have proven
very successful in describing meson phenomenology up to 
$\sim 1\, {\rm GeV}$ \cite{BKY88}.  When trying to include
$0^-$--resonances as elementary fields, however, there is no simple
principle to restrict the form of interactions --- the $\pi'$ has the
same spin--parity quantum numbers as the normal pion
itself. Nevertheless, the normal pion field must decouple from the
``hard'' degrees of freedom in the chiral limit in order to describe
a Goldstone boson.  At the same time, the $\pi'$ contributes to the
axial vector hadronic current, and is thus itself affected by chiral
symmetry: when the axial current is conserved (chiral limit), one
expects the $\pi'$ weak decay constant to vanish \cite{dominguez}. 
This shows that chiral symmetry plays an important role in the 
description of the $\pi'$.
\par
A useful guideline in the construction of effective meson Lagrangians
is the Nambu--Jona-Lasinio (NJL) model, which describes the
spontaneous breaking of chiral symmetry at quark level using a
four--fermion interaction \cite{volkov_83,ebert_86}. The bosonization
of this model and the derivative expansion of the resulting fermion
determinant reproduce the Lagrangian of the linear sigma model, which
embodies the physics of soft pions, as well as higher--derivative
terms.  With appropriate couplings the model allows to derive also a
Lagrangian for vector mesons. This not only gives the correct
structure of the terms of the Lagrangian as required by chiral
symmetry, one also obtains quantitative predictions for the
coefficients, such as $f_\pi$, $g_\rho$, {\em etc.}, which are in
good agreement with phenomenology. One may therefore hope that a
suitable generalization of the NJL--model may provide a means for
deriving an effective Lagrangian including also the $\pi'$ meson.
\par
When extending the NJL model to describe radial excitations of
mesons, one has to introduce non-local (finite--range) four--fermion
interactions.  Many non-local generalizations of the NJL model have
been proposed, using either covariant--euclidean \cite{roberts_88} or
instantaneous (potential--type) \cite{leyaouanc_84,pervushin_90}
effective quark interactions.  These models generally require bilocal
meson fields for bosonization, which makes it difficult to perform a
consistent derivative expansion leading to an effective
Lagrangian\footnote{An exception to this are heavy--light mesons, in
which the heavy quark can be taken as static. An effective Lagrangian
for excited heavy mesons was derived from bilocal fields in
\cite{nowak_93}.}.  A simple alternative is the use of separable
quark interactions.  There are a number of advantages of working with
such a scheme.  First, separable interactions can be bosonized by
introducing local meson fields, just as the usual
NJL--model\footnote{This fact is also important in applications of
this model to finite temperature \cite{schmidt}.}. One can thus
derive an effective meson Lagrangian directly in terms of local
fields and their derivatives. Second, separable interactions allow
one to introduce a limited number of excited states and only in a
given channel. An interesting method for describing excited meson
states in this approximation was proposed in \cite{andrianov_93}.
Furthermore, the separable interaction can be defined in Minkowski
space in a 3--dimensional (yet covariant) way, with form factors
depending only on the part of the quark--antiquark relative momentum
transverse to the meson momentum \cite{pervushin_90,kalinovsky_89}.
This is essential for a correct description of excited states, since
it ensures the absence of spurious relative--time excitations
\cite{feynman_71}.  Finally, as we shall see, the form factors
defining the separable interaction can be chosen in such a way that
the gap equation of the generalized NJL--model coincides with the one
of the usual NJL--model, which has as solution a constant
(momentum--independent) dynamical quark mass. Thus, in this approach
it is possible to describe radially excited mesons above the usual
NJL vacuum. Aside from the technical simplification the latter means
that the separable generalization contains all the successful
quantitative results of the usual NJL model.
\par
In this paper, we derive an effective chiral Lagrangian describing
$\pi$ and $\pi'$ mesons from a generalized NJL--model with separable
interactions. In section 2, we introduce the effective quark
interaction in the separable approximation and describe its
bosonization.  We discuss the choice of form factors necessary to
describe excited states. In section 3, we solve the gap equation
defining the vacuum, derive the effective Lagrangian of the $0^-$
meson fields, and perform the diagonalization leading to the physical
$\pi$ and $\pi'$ states. The effective Lagrangian describes the
vanishing of the $\pi$ mass (decoupling of the Goldstone boson) in
the chiral limit, while the $\pi'$ remains massive.  In section 4, we
derive the axial vector current of the effective Lagrangian using the
Gell-Mann--Levy method and obtain a generalization of the PCAC
formula which includes the contribution of the $\pi'$ to the axial
current. The leptonic decay constants of the $\pi$ and $\pi'$ mesons,
$f_\pi$ and $f_\pi'$, are discussed in section 5.  It is shown that
$f_{\pi'}$ vanishes in the chiral limit as expected.  In section 6,
we fix the parameters of the model and evaluate the ratio 
$f_{\pi'} / f_\pi$ as a function of the $\pi'$ mass.
\par
We stress that we are using the NJL model to derive an effective
lagrangian description of excited mesons, in which the coupling
constants and fields are defined at zero 4--momentum (derivative
expansion), not an on--shell description of bound states. For this
reason the lack of quark confinement of this model is not an issue
here. While there is no {\em a priori} proof of the quantitative
reliability of an effective lagrangian description of excited states,
this approach allows us to investigate a number of principal aspects
in a simple way. Moreover, the fact that effective lagrangians work
well for vector mesons gives some grounds to hope that such an
approach may not be unreasonable for excited states.
\section{Nambu--Jona-Lasinio model with separable interactions}
In the usual NJL model, the spontaneous breaking of chiral symmetry
is described by a local (current--current) effective quark
interaction. The model is defined by the action
\be
S [\bar\psi , \psi ] &=& 
\int d^4 x \, \bar\psi (x) \left( i \partialslash - m^0 \right)
\psi (x) \; + \; S_{\rm int} ,
\label{S_NJL} \\
S_{\rm int} &=& g \int d^4 x \left[ j_\sigma (x) j_\sigma (x) + 
j_\pi^a (x) j_\pi^a (x) \right] ,
\label{S_int}
\ee
where $j_{\sigma , \pi} (x)$ denote, respectively, the 
scalar--isoscalar and pseudoscalar--isovector currents of the 
quark field ($SU(2)$--flavor),
\be
j_\sigma (x) &=& \bar\psi (x) \psi (x), \hspace{2cm}
j_\pi^a (x) \;\; = \;\; \bar\psi (x) i\gamma_5 \tau^a \psi (x) .
\label{j_def}
\ee
The model can be bosonized in the standard way by representing the
4--fermion interaction as a Gaussian functional integral over scalar
and pseudoscalar meson fields \cite{volkov_83,ebert_86}. Since the
interaction, eq.(\ref{S_int}), has the form of a product of two local
currents, the bosonization is achieved through local meson
fields. The effective meson action, which is obtained by integration
over the quark fields, is thus expressed in terms of local meson
fields. By expanding the quark determinant in derivatives of the
local meson fields one then derives the chiral meson Lagrangian.
\par
The NJL interaction, eq.(\ref{S_int}), describes only ground--state
mesons. To include excited states one has to introduce effective
quark interactions with a finite range.  In general, such
interactions require bilocal meson fields for bosonization
\cite{roberts_88,pervushin_90}. A possibility to avoid this
complication is the use of a separable interaction, which is still of
current--current form, eq.(\ref{S_int}), but allows for non-local
vertices (form factors) in the definition of the quark currents,
eq.(\ref{j_def}),
\be
\tilde{S}_{\rm int} &=& g \int d^4 x \sum_{i = 1}^N
\left[ j_{\sigma , i} (x) j_{\sigma , i} (x) 
+ j_{\pi , i}^{a} (x) j_{\pi , i}^{a} (x) \right] , 
\label{int_sep}
\\
j_{\sigma , i} (x) &=& \int d^4 x_1 \int d^4 x_2 \; 
\bar\psi (x_1 ) F_{\sigma , i} (x; x_1, x_2 ) \psi (x_2 ), 
\label{j_sigma} \\
j_{\pi , i}^{a} (x) &=& \int d^4 x_1 \int d^4 x_2 \; 
\bar\psi (x_1 ) F_{\pi , i}^{a} (x; x_1, x_2 ) \psi (x_2 ) . 
\label{j_pi} 
\ee
Here, $F_{\sigma , i}(x; x_1, x_2 ), F_{\pi , i}^{a}(x; x_1, x_2 ), 
\, i = 1, \ldots N$, denote a set of non-local scalar and
pseudoscalar fermion vertices (in general momentum-- and
spin--dependent), which will be specified below. Upon bosonization
eq.(\ref{int_sep}) leads to an action
\be
\lefteqn{
S_{\rm bos}[\bar\psi , \psi; \sigma_1 , \pi_1 , \ldots 
\sigma_N , \pi_N ]  } && \nonumber \\
&=& \int d^4 x_1 \int d^4 x_2 \;
\bar\psi (x_1 ) \left[ \left( i \partialslash_{x_2} - m^0 \right) 
\delta (x_1 - x_2 ) \rule{0cm}{1.5em}
\right. \nonumber \\
&& \left. \rule{0cm}{1.5em}
+ \int d^4 x  \sum_{i = 1}^N \left( \sigma_i (x)
F_{\sigma , i} (x; x_1, x_2 ) + \pi_i^a (x) 
F_{\pi , i}^a (x; x_1, x_2 )
\right) \right] \psi (x_2 ) \nonumber \\
 &-& \frac{1}{2g} \int\! d^4 x \sum_{i = 1}^{N}
\left( \sigma_i^2 (x) + \pi_i^{a\, 2} (x) \right) .
\label{S_sep}
\ee
It describes a system of local meson fields, 
$\sigma_i (x), \pi_i^a (x),\, i = 1, \ldots N$, which interact with
the quarks through non-local vertices. We emphasize that these fields
are not yet to be associated with physical particles ($\sigma ,
\sigma', \ldots, \pi, \pi' , \ldots$); physical fields will be
obtained after determining the vacuum and diagonalizing the effective
meson action.
\par
To define the vertices of eq.(\ref{j_sigma}, \ref{j_pi}) we pass to
the momentum representation. Because of translational invariance, the
vertices can be represented as
\be
\lefteqn{ F_{\sigma , i} (x; x_1, x_2 ) } && \nonumber \\
&=& \fourint{P} \fourint{k}
\exp i \left[ \half (P + k) \cdot (x - x_1 )
+ \half (P - k) \cdot (x - x_2 ) \right]
F_{\sigma , i} (k | P) , \nonumber
\\
\ee
and similarly for $F_{\pi , i}^{a} (x; x_1, x_2 )$. Here $k$ and $P$
denote, respectively, the relative and total momentum of the
quark--antiquark pair. We take the vertices to depend only on the
component of the relative momentum transverse to the total momentum,
\be
F_{\sigma , i} (k | P) &\equiv&
F_{\sigma , i} (k_\perp | P), \hspace{1cm} \mbox{\it etc.}, 
\hspace{2cm}
k_\perp \; \equiv \; k - \frac{P\cdot k}{P^2} P.
\label{markov_yukawa}
\ee
Here, $P$ is assumed to be time-like, $P^2 > 0$.
Eq.(\ref{markov_yukawa}) is the covariant generalization of the
condition that the quark--meson interaction be instantaneuos in the
rest frame of the meson ({\em i.e.}, the frame in which 
${\bf P} = 0$). Eq.(\ref{markov_yukawa}) ensures the absence of
spurious relative--time excitations and thus leads to a consistent
description of excited states\footnote{In bilocal field theory, this
requirement is usually imposed in the form of the so--called
Markov--Yukawa condition of covariant instanteneity of the bound
state amplitude \cite{pervushin_90}. An interaction of transverse
form, eq.(\ref{markov_yukawa}), automatically leads to meson
amplitudes satisfying the Markov--Yukawa condition.}
\cite{feynman_71}.  In particular, this framework allows us to use
3--dimensional ``excited state'' wave functions to model the form
factors for radially excited mesons.
\par
The simplest chirally invariant interaction describing scalar and
pseudoscalar mesons is defined by the spin--independent vertices $1$
and $i\gamma_5 \tau^a$, respectively. We want to include ground state
mesons and their first radial excitation ($N = 2$), and therefore
take
\be
\left.
\ba{r}
F_{\sigma , 1} (k_\perp | P)  \\ 
F_{\pi , 1}^a (k_\perp | P) 
\ea \right\} 
&=& 
\left\{
\ba{r}
1 \\              
i \gamma_5 \tau^a 
\ea \right\}
\times \Theta (\Lambda_3 - | k_\perp | ) 
\label{F_1}
\\
\left.
\ba{r}
F_{\sigma , 2} (k_\perp | P)  \\ 
F_{\pi , 2}^{a} (k_\perp | P) 
\ea \right\} 
&=& 
\left\{
\ba{r}
1 \\              
i \gamma_5 \tau^a 
\ea \right\}
\times \Theta (\Lambda_3 - | k_\perp | ) \, f(k_\perp )
\label{F_2}, 
\\
f(k_\perp ) &=& c (1 + d \, | k_\perp |^2 ) , 
\hspace{2cm} | k_\perp | \;\; \equiv \;\; \sqrt{-k_\perp^2}.
\label{f}
\ee
The step function, $\Theta (\Lambda_3 - | k_\perp | )$, is nothing
but the covariant generalization of the usual 3--momentum cutoff of
the NJL model in the meson rest frame \cite{pervushin_90}. The form
factor $f(k_\perp )$ has for $d < -\Lambda_3^{-2}$ the form of an
excited state wave function, with a node in the interval 
$0 < | k_\perp | < \Lambda_3$.  Eqs.(\ref{F_1}, \ref{F_2}, \ref{f})
are the first two terms in a series of polynomials in $k_\perp^2$;
inclusion of higher excited states would require polynomials of
higher degree.  Note that the normalization of the form factor
$f(k_\perp )$, the constant $c$, determines the overall strength of
the coupling of the $\sigma_2$ and $\pi_2$ fields to the quarks
relative to the usual NJL--coupling of $\pi_1$ and $\sigma_1$.
\par
We remark that the most general vertex could also include
spin--dependent structures, $\Pslash$ and $\gamma_5\Pslash$, which in
the terminology of the NJL model correspond to the induced vector and
axial vector component of the $\sigma$ and $\pi$ ($\sigma$--$\rho$
and $\pi$--$A_1$ mixing), respectively. These structures should be
considered if vector mesons are included. Furthermore, there could be
structures $\kslash_\perp , \Pslash \kslash_\perp$ and
$\gamma_5\kslash_\perp , \gamma_5\Pslash \kslash_\perp$,
respectively, which describe bound states with orbital angular
momentum $L = 1$.  We shall not consider these components here.
\par
With the form factors defined by eqs.(\ref{F_1}, \ref{F_2}, \ref{f}),
the bosonized action, eq.(\ref{S_sep}), in momentum representation
takes the form
\be
\lefteqn{
S_{\rm bos}[\bar\psi , \psi; \sigma_1 , \pi_1 , 
\sigma_2 , \pi_2 ]  } && 
\nonumber \\
&=& \fourint{k} \bar\psi (k) \left( \kslash - m^0 \right) \psi (k) 
\nonumber \\
&+& \fourint{P} \kint \bar\psi (k + \half P) 
\left[ \sigma_1 (P) + i\gamma_5 \tau^a \pi_1^a (P) 
 + \left( \sigma_2 (P) + i\gamma_5 \tau^a \pi_2^a (P) \right) 
f(k_\perp ) \right] 
\nonumber \\
&& \times \psi (k - \half P) 
\nonumber \\
&-& \frac{1}{2g} \fourint{P} \sum_{i = 1}^2
\left( \sigma_i (-P) \sigma_i (P) + \pi_i^a (-P) \pi_i^a (P) 
\right) .
\label{S_momentum}
\ee
Here it is understood that a cutoff in 3--dimensional transverse
momentum is applied to the $k$--integral, as defined by the step
function of eqs.(\ref{F_1}, \ref{F_2}).
\section{ Effective Lagrangian for $\pi$ and $\pi'$ mesons}
We now want to derive the effective Lagrangian describing physical
$\pi$ and $\pi'$ mesons. Integrating over the fermion fields in
eq.(\ref{S_momentum}), one obtains the effective action of the
$\sigma_1 , \pi_1$-- and $\sigma_2 , \pi_2$--fields,
\be
W[\sigma_1 , \pi_1 , \sigma_2 , \pi_2] &=& 
-\frac{1}{2 g} \int  (\sigma_1^2 + \pi_1^{a\, 2} +
\sigma_2^2 + \pi_2^{a\, 2} ) \nonumber \\
&-& i N_c \; {\rm Tr}\, \log \left[ 
i \partialslash - m^0 + \sigma_1 + i \gamma_5 \tau^a \pi_1^a
+(\sigma_2 + i \gamma_5 \tau^a \pi_2^a) f \right].
\label{W}
\ee
This expression is understood as a shorthand notation for expanding
in the meson fields.  In particular, we want to derive the free part
of the effective action for the $\pi_1$-- and $\pi_2$--fields,
\be
W &=& W^{(0)} + W^{(2)}, \\
W^{(2)} &=& \half \int\frac{d^4 P}{(2\pi )^4} 
\sum_{i, j = 1}^{2} \pi_i^a (P) K_{ij}^{ab} (P) \pi_j^b (P) ,
\label{W_2}
\ee
where it is understood that we restrict ourselves to timelike
momenta, $P^2 > 0$.  Before expanding in the $\pi_1$-- and
$\pi_2$--fields, we must determine the vacuum, {\em i.e.}, the mean
scalar field, which arises in the dynamical breaking of chiral
symmetry.  The mean--field approximation corresponds to the leading
order of the $1/N_c$--expansion. The mean field is determined by the
set of equations
\be
\frac{\delta W}{\delta\sigma_1} &=& - i N_c \; {\rm tr} \kint
\frac{1}{\rlap/k - m^0 + \sigma_1 + \sigma_2 f(k_\perp )}
- \frac{\sigma_1}{g} \; = \; 0 , \label{gap_1}\\
\frac{\delta W}{\delta\sigma_2} &=& - i N_c \; {\rm tr} \kint
\frac{f(k_\perp )}{\rlap/k - m^0 + \sigma_1 + \sigma_2 f(k_\perp )}
- \frac{\sigma_2}{g} \; = \; 0 . \label{gap_2}
\ee
Due to the transverse definition of the interaction,
eq.(\ref{markov_yukawa}), the mean field inside a meson depends in a
trivial way on the direction of the meson 4--momentum, $P$. In the
following we consider these equations in the rest frame where
${\bf P} = 0, k_\perp = (0, \bfk )$ and $\Lambda_3$ is the usual
3--momentum cutoff.
\par
In general, the solution of eqs.(\ref{gap_1}, \ref{gap_2}) would have
$\sigma_2\neq 0$, in which case the dynamically generated quark mass,
$-\sigma_1 - \sigma_2 f(\bfk ) + m^0$, becomes momentum--dependent.
However, if we choose the form factor, $f(\bfk )$, such that
\be
-4 m I_1^f &\equiv&
 - i N_c \; {\rm tr} \kint
\frac{f(\bfk )}{\rlap/k - m} \;\; =\;\;
i 4 N_{\rm c} m \kint \frac{f(\bfk )}{m^2 - k^2} \;\; = \;\; 0 , 
\label{cond} \\
m &\equiv& - \sigma_1 + m^0 ,
\nonumber
\ee
eqs.(\ref{gap_1}, \ref{gap_2}) admit a solution with $\sigma_2 = 0$
and thus with a constant quark mass, $m = -\sigma_1 + m^0$. In this
case, eq.(\ref{gap_1}) reduces to the usual gap equation of the NJL
model,
\be
- 8 m I_1
&\equiv& -8 m i N_{\rm c} \kint \frac{1}{k^2 - m^2} 
\; = \; \frac{m^0 - m}{g}.
\label{gap_njl}
\ee
Obviously, condition eq.(\ref{cond}) can be fulfilled by choosing an
appropriate value of the parameter $d$ defining the ``excited state''
form factor, eq.(\ref{f}), for given values of $\Lambda_3$ and
$m$. Eq.(\ref{cond}) expresses the invariance of the usual NJL
vacuum, $\sigma_1 = {\rm const.}$, with respect to variations in the
direction of $\sigma_2$. In the following, we shall consider the
vacuum as defined by eqs.(\ref{cond}, \ref{gap_njl}), {\em i.e.}, we
work with the usual NJL vacuum. We emphasize that this choice is a
matter of convenience, not of principle. The qualitative results
below could equivalently be obtained with a different choice of form
factor; however, in this case one would have to re-derive all vacuum
and ground--state meson properties with the momentum--dependent quark
mass. Preserving the NJL vacuum makes formulas below much more
transparent and allows to carry over the parameters fixed in the old
NJL model.
\par
With the mean field determined by eqs.(\ref{cond}, \ref{gap_njl}), we
now expand the action to quadratic order in the fields $\pi_1$ and
$\pi_2$.  The quadratic form $K_{ij}^{ab} (P)$, eq.(\ref{W_2}), is
obtained as
\be
K_{ij}^{ab} (P) &\equiv& \delta^{ab} K_{ij} (P) , \nonumber \\
K_{ij} (P) &=& -i N_{\rm c} \; {\rm tr}\, \kint \left[
\frac{1}{\kslash + \half\Pslash - m}
i\gamma_5 f_i
\frac{1}{\kslash - \half\Pslash - m} i \gamma_5 f_j
\right]  - \delta_{ij} \frac{1}{g} , \nonumber
\\
f_1 &\equiv& 1, \hspace{2em} f_2 \;\; \equiv \;\; f(k_\perp ).
\label{K_full}
\ee
A graphical representation of the loop integrals in eq.(\ref{K_full})
is given in fig.1.  The integral is evaluated by expanding in the
meson field momentum, $P$. To order $P^2$, one obtains
\be
K_{11}(P) &=& Z_1 (P^2 - m_1^2 ), 
\hspace{2em} K_{22}(P) \;\; = \;\; Z_2 (P^2 - m_2^2 ) \nonumber \\
K_{12}(P) &=& K_{21}(P) \;\; = \;\;
\sqrt{Z_1 Z_2} \, \Gamma P^2 
\label{K_matrix}
\ee
where
\be
Z_1 &=& 4 I_2, \hspace{2em} Z_2 \; = \; 4 I_2^{ff}, 
\label{I_12} \\
m_1^2 &=& Z_1^{-1}(-8I_1 + g^{-1}) \; = \; \frac{m^0}{Z_1 g m} , 
\label{m_1} \\
m_2^2 &=& Z_2^{-1}(-8I_1^{ff} + g^{-1}) , 
\label{m_2} \\
\Gamma &=& \frac{4}{\sqrt{Z_1 Z_2}} I_2^f .
\label{gamma}
\ee
Here, $I_n, I_n^f$ and $I_n^{ff}$ denote the usual loop integrals
arising in the momentum expansion of the NJL quark determinant, but
now with zero, one or two factors $f(k_\perp )$, eq.(\ref{f}), in the
numerator. We may evaluate them in the rest frame, 
$k_\perp = (0, \bfk )$,
\be
I_n^{f..f} &\equiv& -i N_{\rm c} 
\kint \frac{f(\bfk )..f(\bfk )}{(m^2 - k^2)^n}.
\label{I_n}
\ee
The evaluation of these integrals with a 3--momentum cutoff is
described {\em e.g.}\ in ref.\cite{Ebert_93}. The integral over $k_0$
is performed by contour integration, and the remaining 3--dimensional
integral regularized by the cutoff. Only the divergent parts are
kept; all finite parts are dropped. We point out that the momentum
expansion of the quark loop integrals, eq.(\ref{K_full}), is an
essential part of this approach.  The NJL--model is understood here
as a model only for the lowest coefficients of the momentum expansion
of the quark loop, not its full momentum dependence (singularities
{\em etc.}).
\par
Note that a mixing between the $\pi_1$-- and $\pi_2$--fields occurs
only in the kinetic (${\cal O}(P^2 )$--) terms of
eq.(\ref{K_matrix}), but not in the mass terms. This is a direct
consequence of the definition of the vacuum by eqs.(\ref{cond},
\ref{gap_njl}), which ensures that the quark loop with one form
factor has no $P^2$--independent part. The ``softness'' of the
$\pi_1$--$\pi_2$ mixing causes the $\pi_1$--field to decouple at
$P^2\rightarrow 0$. This property is crucial for the appearance of a
Goldstone boson in the chiral limit.
\par
To determine the physical $\pi$-- and $\pi'$--meson states we have to
diagonalize the quadratic part of the action, eq.(\ref{W_2}). If one
knew the full momentum dependence of the quadratic form,
eq.(\ref{K_matrix}), the masses of the physical states would be given
as the zeros of the determinant of the quadratic form,
\be
\det K_{ij} (P^2 ) &=& 0, \hspace{2cm} P^2 \;\; = \;\; m_\pi^2 , \;
m_{\pi'}^2  .
\label{determinant}
\ee
This would be equivalent to the usual Bethe--Salpeter (on--shell)
description of bound states: the matrix $K_{ij}(P^2)$ is diagonalized
independently on the respective mass shells, 
$P^2 = m_\pi^2 , m_{\pi'}^2$ \cite{leyaouanc_84,weiss_93,gross}.  In
our approach, however, we know the quadratic form,
eq.(\ref{K_matrix}), only as an expansion in $P^2$ at $P^2 = 0$. It
is clear that a determination of the masses according to
eq.(\ref{determinant}) would be incompatible with the momentum
expansion, as the determinant involves ${\cal O}(P^4 )$--terms which
are neglected in eq.(\ref{K_matrix}).  To be consistent with the
$P^2$--expansion we must diagonalize the kinetic term and the mass
term in eq.(\ref{W_2}) simultaneously, with a $P^2$--independent
transformation of the fields. Let us write eq.(\ref{K_matrix}) in
matrix form
\be
K_{ij}(P^2 ) &=& \left(\ba{ll}
Z_1 & \sqrt{Z_1 Z_2} \, \Gamma \\
\sqrt{Z_1 Z_2} \, \Gamma & Z_2 \ea\right) P^2 \;\; - \;\;
\left(\ba{ll}
Z_1 m_1^2 &  \\ & Z_2 m_2^2 \ea\right) .
\ee
The transformation which diagonalizes both matrices here separately
is given by
\be
\ba{lcrcr}
\sqrt{Z_1} \pi_1^a &=& 
{\displaystyle \frac{\cos\phi}{\sqrt{{Z_\pi}}} \pi^a }
&+& {\displaystyle \frac{m_2}{m_1} \frac{\sin\phi}{\sqrt{Z_{\pi'}}} 
\pi^{\prime\, a}} , 
\\[.1cm] 
\sqrt{Z_2} \pi_2^a &=& {\displaystyle
\frac{m_1}{m_2}  \frac{\sin\phi}{\sqrt{Z_\pi}} \pi^a }
&-& {\displaystyle \frac{\cos\phi}{\sqrt{Z_{\pi'}}} 
\pi^{\prime\, a} } ,
\ea
\label{transform}
\ee
where
\be
\tan 2\phi &=& 2 \Gamma \frac{m_1}{m_2} 
\left( 1 - \frac{m_1^2}{m_2^2} \right)^{-1} , 
\label{alpha} \\
Z_\pi &=& \cos^2\phi + \frac{m_1^2}{m_2^2} \sin^2\phi
+ 2 \Gamma \frac{m_1}{m_2} \cos\phi\sin\phi , \\
Z_{\pi'} &=& \cos^2\phi + \frac{m_2^2}{m_1^2} \sin^2\phi
- 2 \Gamma \frac{m_2}{m_1} \cos\phi\sin\phi .
\ee
In terms of the new fields, $\pi , \pi'$, the quadratic part of the
action, eq.(\ref{W_2}), reads
\be 
W^{(2)} &=& \half \fourint{P} \left[ 
\pi^a (-P) (P^2 - m_\pi^2) \pi^a (P)
+ \pi^{\prime\, a} (-P) (P^2 - m_{\pi'}^2) \pi^{\prime\, a} (P) 
\right] .
\label{S_2_phys}
\ee
Here,
\be
m_\pi^2 &=& \frac{m_1^2}{Z_\pi}, \hspace{2em}
m_{\pi'}^2  \;\; = \;\; \frac{m_2^2}{Z_{\pi'}} .
\label{mpi_phys}
\ee
The fields $\pi$ and $\pi'$ can thus be associated with physical
particles.
\par
Let us now consider the chiral limit, {\em i.e.}, vanishing current
quark mass, $m^0 \rightarrow 0$. From eqs.(\ref{I_12}--\ref{gamma})
we see that this is equivalent to taking $m_1^2 \rightarrow 0$.
(Here and in the following, when discussing the dependence of
quantities on the current quark mass, $m^0$, we keep the constituent
quark mass fixed and assume the coupling constant, $g$, to be changed
in accordance with $m^0$, such that the gap equation,
eq.(\ref{gap_njl}) remains fulfilled exactly. In this way, the loop
integrals and eq.(\ref{cond}) remain unaffected by changes of the
current quark mass.)  Expanding eqs.(\ref{mpi_phys}) in 
$m_1^2 \propto m^0$, one finds
\be
m_\pi^2 &=& m_1^2 \; + \; {\cal O}(m_1^4 ), 
\label{mpi_chiral}
\\
m_{\pi'}^2 &=& \frac{m_2^2}{1 - \Gamma^2} 
\left[ 1 \; + \; \Gamma^2 \frac{m_1^2}{m_2^2}
 \; + \; {\cal O}(m_1^4 ) \right] .
\label{mpip_chiral}
\ee
Thus, in the chiral limit the effective Lagrangian
eq.(\ref{S_2_phys}) indeed describes a massless Goldstone pion,
$\pi$, and a massive particle, $\pi'$.  Furthermore, in the chiral
limit the transformation of the fields, eq.(\ref{transform}), becomes
\be
\ba{lcrcr}
\sqrt{Z_1} \pi_1^a &=& 
{\displaystyle \left( 1 - \Gamma^2 \frac{m_1^2}{m_2^2} \right) 
\pi^a}
&+& {\displaystyle \frac{\Gamma}{\sqrt{1 - \Gamma^2}} \left(
1 + (1 - \Gamma^2 ) \frac{m_1^2}{m_2^2} \right) 
\pi^{\prime\, a}} , \\
\sqrt{Z_2} \pi_2^a &=& 
{\displaystyle \Gamma \frac{m_1^2}{m_2^2} \pi^a }
&-& {\displaystyle \frac{1}{\sqrt{1 - \Gamma^2}} 
\pi^{\prime\, a} } . 
\ea
\label{transform_chiral}
\ee
At $m_1^2 = 0$ one observes that $\pi$ has only a component along
$\pi_1$. This is a consequence of the fact that the $\pi_1$--$\pi_2$
coupling in the original Lagrangian, eq.(\ref{K_matrix}), is of order
$P^2$.  We remark that, although we have chosen to work with a
particular choice of excited--state form factor, eq.(\ref{cond}), the
occurence of a Goldstone boson in the chiral limit in eq.(\ref{W}) is
general and does not depend on this choice.  This may easily be
established using the general gap equations, eqs.(\ref{gap_1},
\ref{gap_2}), together with eq.(\ref{K_full}).
\section{The axial current}
In order to describe the leptonic decays of the $\pi$ and $\pi'$
mesons we need the axial current operator.  Since our effective
action contains besides the pion a field describing an ``excited
state'' with the same quantum numbers, it is clear that the axial
current of our model is, in general, not carried exclusively by the
$\pi$ field, and is thus not given by the standard PCAC formula.
Thus, we must determine the conserved axial current of our model,
including the contribution of the $\pi'$, from first principles.
\par
In general, the construction of the conserved current in a theory
with non-local (momentum--dependent) interactions is a difficult
task. This problem has been studied extensively in the framework of
the Bethe--Salpeter equation \cite{riska_87} and various
3--dimensional reductions of it such as the quasipotential and the
on--shell reduction \cite{gross_current}. In these approaches the
derivation of the current is achieved by ``gauging'' all possible
momentum dependence of the interaction through minimal substitution,
a rather cumbersome procedure in practice. In contrast, in a
Lagrangian field theory a simple method exists to derive conserved
currents, the so--called Gell--Mann and Levy method \cite{gell-mann},
which is based on Noether's theorem.  In this approach the current is
obtained as the variation of the lagrgangian with respect to the
derivative of a space--time dependent symmetry transformation of the
fields. We now show that a suitable generalization of this technique
can be employed to derive the conserved axial current of our model
with quark--meson form factors depending on transverse momentum.
\par
To derive the axial current we start at quark level.  The isovector
axial current is the Noether current corresponding to infinitesimal
chiral rotations of the quark fields,
\be
\psi (x) &\rightarrow& \left( 1 - i \varepsilon^a \half \tau^a 
\gamma_5 \right) \psi (x) .
\label{chiral_psi}
\ee
Following the usual procedure, we consider the parameter of this
transformation to be space--time dependent, 
$\varepsilon^a \equiv \varepsilon^a (x)$. However, this dependence
need not be completely arbitrary. To describe the decays of $\pi$ and
$\pi'$ mesons, it is sufficient to know the component of the axial
current parallel to the meson 4--momentum, $P$. It is easy to see
that this component is obtained from chiral rotations whose parameter
depends only on the longitudinal part of the coordinate
\be
\varepsilon^a (x) &\rightarrow& \varepsilon^a (x_{\vert\vert}),
\hspace{2cm}
x_{\vert\vert} \;\; \equiv \;\; \frac{x \cdot P}{\sqrt{P^2}}, 
\label{eps_x}
\ee
since $\partial_\mu \varepsilon^a (x_{\vert\vert}) \propto P_\mu$.
In other words, transformations of the form eq.(\ref{eps_x}) describe
a transfer of longitudinal momentum to the meson, but not of
transverse momentum. This has the important consequence that the
chiral transformation does not change the direction of transversality
of the meson--quark interaction, {\em cf.}\ eq.(\ref{markov_yukawa}).
When passing to the bosonized representation, eq.(\ref{S_sep}), the
transformation of the $\pi_1 , \sigma_1$-- and 
$\pi_2 , \sigma_2$--fields induced by eqs.(\ref{chiral_psi},
\ref{eps_x}) is therefore of the form
\be
\ba{lclcl}
\pi^a_i (x) &\rightarrow& \pi^a_i (x) &+& 
\varepsilon^a (x_{\vert\vert}) \, \sigma_i (x) , \\
\sigma_i (x) &\rightarrow& \sigma_i (x) &-& 
\varepsilon^a (x_{\vert\vert}) \, \pi^a_i (x) .
\ea
\hspace{1.5cm} (i = 1, 2)
\label{chiral_pi}
\ee
This follows from the fact that, for fixed direction of $P$, the
vertex eq.(\ref{markov_yukawa}) describes an instantaneous
interaction in $x_{\vert\vert}$. Thus, the special chiral rotation
eq.(\ref{eps_x}) does not mix the components of the meson fields
coupling to the quarks with different form factors.
\par
With the transformation of the chiral fields given by
eqs.(\ref{chiral_pi}), the construction of the axial current proceeds
exactly as in the usual linear sigma model. We write the variation of
the effective action, eq.(\ref{W}), in momentum representation,
\be
\delta W &=& \fourint{Q} \varepsilon^a (Q) D^a (Q) ,
\ee
where $\varepsilon^a (Q) = \tilde\varepsilon^a (Q_{\vert\vert})
\delta^{(3)} (Q_\perp )$ is the Fourier transform of the
transformation eq.(\ref{eps_x}) and $D^a (Q)$ is a function of the
fields $\sigma_i , \pi_i , i = 1, \ldots 2$, given in the form of a
quark loop integral,
\be
D^a (Q)
&=& -i N_c \; {\rm tr} \fourint{k} 
\left[ \frac{1}{\kslash - m}\delta^{ab}
+ \frac{1}{\kslash - \half \Qslash - m} i\gamma_5 \tau^a
\frac{1}{\kslash + \half \Qslash - m} i\gamma_5 \tau^b 
\sigma_1 \right] 
\nonumber \\
&& \times (\pi_1^b (Q) + f(k_\perp ) \pi_2^b (Q) ) .
\ee
Here we have already used that $\sigma_2 = 0$ in the vacuum,
eq.(\ref{cond}). Expanding now in the momentum $Q$, making use of
eq.(\ref{cond}) and the gap equation, eq.(\ref{gap_njl}), and setting
$\sigma_1 = -m$ (it is sufficient to consider the symmetric limit,
$m^0 = 0$), this becomes
\be
D^a (Q)
&=& -Q^2 m \left[ 4 I_2 \pi_1^a (Q) + 4 I_2^f \pi_2^a (Q) \right] 
\nonumber \\
&=& -Q^2 m \left[ Z_1 \pi_1^a (Q) + \sqrt{Z_1 Z_2} \Gamma
\pi_2^a (Q) \right] . 
\label{D_result}
\ee
The fact that $D^a (Q^2 )$ is proportional to $Q^2$ is a consequence
of the chiral symmetry of the effective action, eq.(\ref{W}). Due to
this property, $D^a (Q^2 )$ may be regarded as the divergence of a
conserved current,
\be
A_\mu^a (Q) &=& Q_\mu m \left[ 
Z_1 \pi_1^a (Q) + \sqrt{Z_1 Z_2} \Gamma \pi_2^a (Q) \right] .
\label{axial_current}
\ee
Eq.(\ref{axial_current}) is the conserved axial current of our model.
It is of the usual ``PCAC'' form, but contains also a contribution of
the $\pi_2$ field. The above derivation was rather formal. However,
the result can be understood in simple terms, as is shown in fig.2:
Both the $\pi_1$ and $\pi_2$--field couple to the local axial current
of the quark field through quark loops; the $\pi_2$--field enters the
loop with a form factor, $f(k_\perp )$. The necessity to pull out a
factor of the meson field momentum (derivative) means that only the
${\cal O}(P^2)$--parts of the loop integrals, $I_2$ and $I_2^f$,
survive, {\em cf.}\ eq.(\ref{I_n}). Chiral symmetry ensures that the
corresponding diagrams for the divergence of the current have no
$P^2$--independent part.
\par
The results of this section are an example for the technical
simplifications of working with separable quark interactions. The
fact that they can be bosonized by local meson fields makes it
possible to apply methods of local field theory, such as Noether's
theorem, to the effective meson action. Furthermore, we note that the
covariant (transverse) definition of the 3--dimensional quark
interaction, eq.(\ref{markov_yukawa}), is crucial for obtaining a
consistent axial current. In particular, with this formulation there
is no ambiguity with different definitions of the pion decay constant
as with non--covariant 3-dimensional interactions
\cite{leyaouanc_84}.
\section{ The weak decay constants of the $\pi$ and $\pi'$ meson}
We now use the axial current derived in the previous section to
evaluate the weak decay constants of the physical $\pi$ and $\pi'$
mesons. They are defined by the matrix element of the divergence of
the axial current between meson states and the vacuum,
\be
\langle 0 | \partial^\mu A_\mu^a | \pi^b \rangle &=&
m_\pi^2 f_\pi \delta^{ab} ,
\label{decay_pi} \\
\langle 0 | \partial^\mu A_\mu^a | \pi^{\prime\, b} \rangle &=&
m_{\pi'}^2 f_{\pi'} \delta^{ab}
\label{decay_pipr} .
\ee
In terms of the physical fields, $\pi$ and $\pi'$, the axial current
takes the form
\be
A_\mu^a &=& P_\mu m \sqrt{Z_1} \left( \pi^a 
\; + \; \Gamma \sqrt{1 - \Gamma^2} \, 
\frac{m_1^2}{m_2^2} \, \pi^{\prime\, a} \right) 
\; + \; {\cal O}(m_1^4 ).
\ee 
Here, we have substituted the transformation of the fields,
eq.(\ref{transform_chiral}), in eq.(\ref{axial_current}).  The decay
constants of the physical $\pi$ and $\pi'$ states are thus given by
\be
f_\pi &=& \sqrt{Z_1} m \; + \; {\cal O}(m_1^4 ) 
\label{fpi} \\
f_{\pi'} &=& \sqrt{Z_1} m \, \Gamma \sqrt{1 - \Gamma^2} \, 
\frac{m_1^2}{m_2^2} \; + \; {\cal O}(m_1^4 ) .
\label{fpip}
\ee
The corrections to $f_\pi$ due to the inclusion of excited states are
of order $m_\pi^4$. Thus, within our accuracy $f_\pi$ is identical to
the value obtained with the usual NJL model, $\sqrt{Z_1} m$, which 
follows from the Goldberger--Treiman relation at quark 
level \cite{volkov_83}.  On the other hand, the $\pi'$ decay constant
vanishes in the chiral limit $m^0 \sim m_1^2 \rightarrow 0$, as
expected. We stress that for this property to hold it is essential to
consider the full axial current, eq.(\ref{axial_current}), including
the contribution of the $\pi_2$--component. As can be seen from
eqs.(\ref{transform_chiral}, \ref{axial_current}), use of the
standard PCAC formula $A_\mu^a \propto \partial_\mu \pi_1^a$ would
lead to a non-vanishing result for $f_{\pi'}$ in the chiral limit.
\par
The ratio of the $\pi'$ to the $\pi$ decay constants can directly be
expressed in terms of the physical $\pi$ and $\pi'$ masses.  From
eqs.(\ref{fpi}, \ref{fpip}) one obtains, using eqs.(\ref{mpi_chiral},
\ref{mpip_chiral}),
\be
\frac{f_{\pi'}}{f_\pi} &=& \Gamma \sqrt{1 - \Gamma^2} \, 
\frac{m_1^2}{m_2^2}
\;\; = \;\; \frac{\Gamma}{\sqrt{1 - \Gamma^2}} \,
\frac{m_\pi^2}{m_{\pi'}^2}
\label{ratio}
\ee
This is precisely the dependence which was derived from current
algebra considerations in a general ``extended PCAC'' framework
\cite{dominguez}.  We note that the same behavior of $f_{\pi'}$ in
the chiral limit is found in models describing chiral symmetry
breaking by non-local interactions \cite{leyaouanc_84,weiss_93}.
\par
The effective Lagrangian illustrates in a compact way the different
consequences of axial current conservation for the pion and its
excited state. Both matrix elements of $\partial_\mu A^\mu$,
eq.(\ref{decay_pi}) and eq.(\ref{decay_pipr}), must vanish for 
$m^0 \rightarrow 0$.  The pion matrix element, eq.(\ref{decay_pi})
does so by $m_\pi^2 \rightarrow 0$, with $f_\pi$ remaining finite,
while for the excited pion matrix element the opposite takes place,
$f_{\pi'} \rightarrow 0$ with $m_{\pi'}$ remaining finite.
\section{ Numerical estimates and conclusions }
We can now estimate numerically the excited pion decay constant,
$f_{\pi'}$, in this model. We take a value of the constituent quark
mass of $m = 300\,{\rm MeV}$ and fix the 3--momentum cutoff at
$\Lambda_3 = 671 \,{\rm MeV}$ by fitting the normal pion decay
constant $f_\pi = 93\, {\rm MeV}$ in the chiral limit, as in the
usual NJL model without excited states, {\em cf.} \cite{Ebert_93}.
With these parameters one obtains the standard value of the quark
condensate, $\langle \bar q q \rangle = - (253 \, {\rm MeV})^3$, and
$g = 0.82\, m^{-2} = 9.1\, {\rm GeV}^{-2}, \; m^0 = 5.1\, {\rm MeV}$.
With the constituent quark mass and cutoff fixed, we can determine
the parameter $d$ of the ``excited--state'' form factor,
eq.(\ref{f}), from the condition eq.(\ref{cond}).  We find 
$d = -1.83 \, \Lambda_3^{-2} = -4.06\, {\rm GeV}^{-2}$, corresponding
to a form factor $f(k_\perp )$ with a radial node in the range 
$0 \leq | k_\perp | \leq \Lambda_3$.  With this value we determine
the $\pi_1$--$\pi_2$ mixing coefficient, $\Gamma$, eq.(\ref{gamma}),
as
\be
\Gamma &=& 0.41.
\ee
Note that $\Gamma$ is independent of the normalization of the form
factor $f(k_\perp )$, eq.(\ref{f}). In fact, the parameter $c$ enters
only the mass of the $\pi'$ meson, {\em cf.}\ eqs.(\ref{m_2},
\ref{mpip_chiral}); we do not need to determine its value since the
result can directly be expressed in terms of $m_{\pi'}$. Thus,
eq.(\ref{ratio}) gives
\be
\frac{f_{\pi'}}{f_\pi} &=& 0.45 \frac{m_\pi^2}{m_{\pi'}^2}
\ee
For the standard value of the $\pi'$ mass, 
$m_{\pi'} = 1300\, {\rm MeV}$, this comes to 
$f_{\pi'} = 0.48 \, {\rm MeV}$, while for a low mass of 
$m_{\pi'} = 750\, {\rm MeV}$ one obtains 
$f_{\pi'} = 1.46 \, {\rm MeV}$.  The excited pion leptonic decay
constant is thus very small, which is a consequence of chiral
symmetry. Note that, as opposed to the qualitative results discussed
above, the numerical values here depend on the choice of form factor,
eq.(\ref{cond}), and should thus be regarded as a rough estimate.
\par
We remark that the numerical values of the ratio $f_{\pi'}/f_\pi$
obtained here are comparable to those found in chirally symmetric
potential models \cite{weiss_93}. However, models describing chiral
symmetry breaking by a vector--type confining potential (linear or
oscillator) usually understimate the normal pion decay constant by an
order of magnitude \cite{leyaouanc_84}. Such models should include a
short--range interaction (NJL--type), which is mostly responsible for
chiral symmetry breaking.
\par
The small value of $f_{\pi'}$ does not imply a small width of the
$\pi'$ resonance, since it can decay hadronically, {\em e.g.}\ into
$3\pi$ or $\rho\pi$. Such hadronic decays can also be investigated in
the chiral Lagrangian framework set up here. This problem, which
requires the evaluation of quark loops with external meson fields of
different 4--momenta, will be left for a future investigation.
\par
In conclusion, we have outlined a simple framework for including
radial excitations in an effective Lagrangian description of
mesons. The Lagrangian obtained by bosonization of an NJL--model with
separable interactions exhibits all qualitative properties expected
on general grounds: a Goldstone pion with a finite decay constant,
and a massive ``excited state'' with vanishing decay constant in the
chiral limit. Our model shows in a simple way how chiral symmetry
protects the pion from modifications by excited states, which in turn
influences the excited states' contribution to the axial
current. These features are general and do not depend on a particular
choice of quark--meson form factor.  Furthermore, they are preserved
if the derivative expansion of the quark loop is carried to higher
orders.
\par
In the investigations described here we have remained strictly within
an effective Lagrangian approach, where the coupling constants and
field transformations are defined at zero momentum. We have no way to
check the quantitative reliability of this approximation for radially
excited states in the region of $\sim 1\,{\rm GeV}$, {\em i.e.}, to
estimate the momentum dependence of the coupling constants, within
the present model. (For a general discussion of the range of
applicability of effective Lagrangians, see \cite{jaffe_92}.)  This
question can be addressed in generalizations of the NJL model with
quark confinement, which in principle allow both a zero--momentum as
well as an on--shell description of bound states. Recently, first
steps were taken to investigate the full momentum dependence of
correlation functions in such an approach \cite{celenza}.
\par
This work was supported partly by the Russian Foundation of
Fundamental Research (Grant N 96.01.01223), by the DFG and by COSY
(J\"ulich).
%

%
%
\newpage
\begin{figure}[h]
\epsfxsize=16cm
\epsfysize=9cm
\centerline{\epsffile{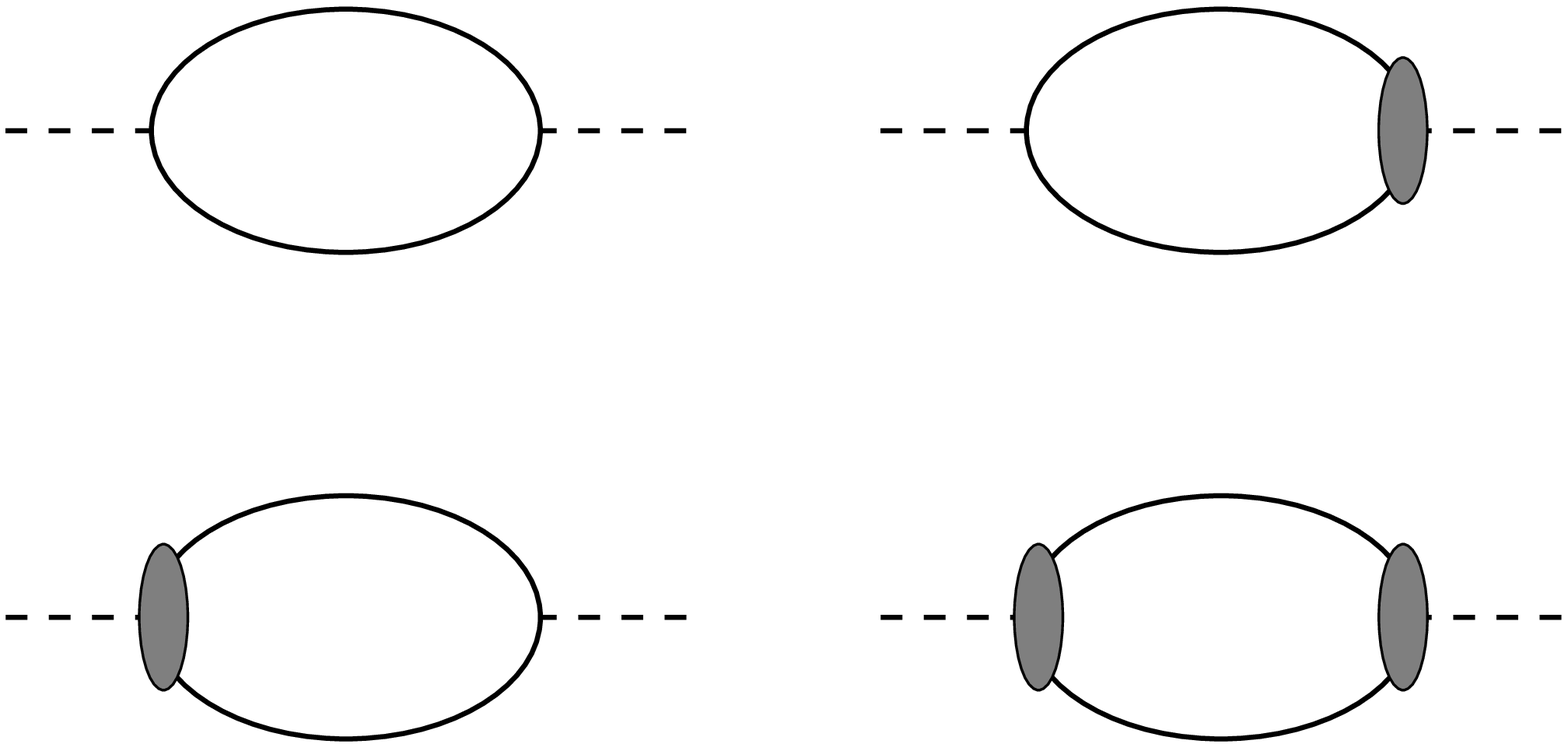}}
\vspace{2cm}
\caption[]{The quark loop contribution to the quadratic form 
$K_{ij}(P)$, eq.(\ref{K_full}), of the effective action of the
$\pi_1$-- and $\pi_2$--fields. The solid lines denote the NJL quark
propagator. The $\pi_1$--field couples to the quarks with a local
vertex, the $\pi_2$--field through the form factor, $f(k_\perp )$,
represented here by a shaded blob.}
\end{figure}
\newpage
\begin{figure}[h]
\centerline{
\mbox{\LARGE\bf $\pi_1 \times \partial_\mu$ \hspace{.1cm}}
\epsfxsize=4.5cm
\epsfysize=2.4cm
\parbox{4.5cm}{\epsffile{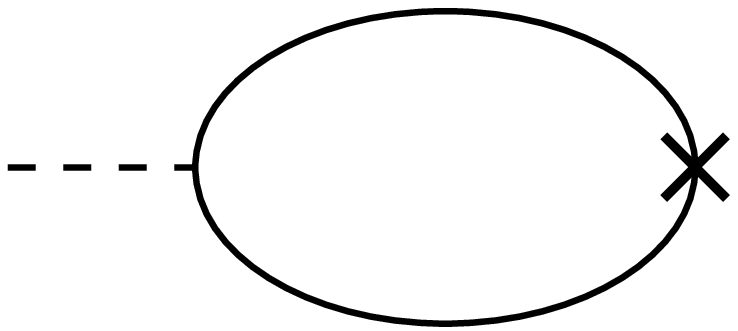}} 
\mbox{\LARGE\bf $\hspace{.5cm} +  \hspace{.5cm}
\pi_2 \times \partial_\mu \hspace{.1cm} $}
\epsfxsize=5cm
\epsfysize=2.4cm
\parbox{4.5cm}{\epsffile{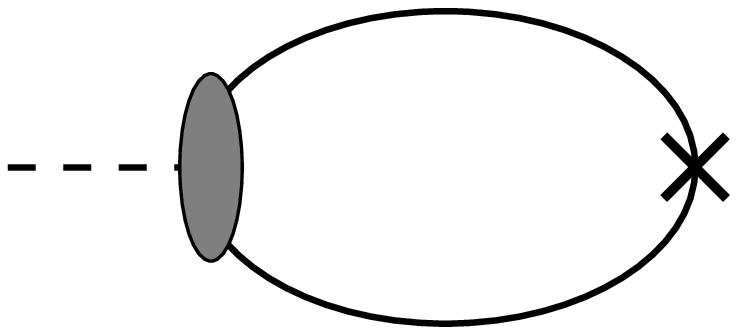}} }
\vspace{2cm}
\caption[]{The axial current of the $\pi_1$-- and $\pi_2$--fields,
eq.(\ref{axial_current}), as it follows from Noether's theorem.  The
cross denotes the local axial current of the quark field, to which
the $\pi_1$-- and $\pi_2$--fields couple through quark
loops. Notation is the same as in fig.1.}
\end{figure}
\end{document}